# A New Multilabel System for Automatic Music Emotion Recognition

Fabio Paolizzo
*Department of Electronic Engineering*
*University of Rome Tor Vergata*
Rome, Italy
fabio.paolizzo@gmail.com

Natalia Pichierri
*Department of Electronic Engineering*
*University of Rome Tor Vergata*
Rome, Italy
pichierrinatalia@gmail.com

Daniele Giardino
*Department of Electronic Engineering*
*University of Rome Tor Vergata*
Rome, Italy
giardino@ing.uniroma2.it

Marco Matta
*Department of Electronic Engineering*
*University of Rome Tor Vergata*
Rome, Italy
matta@ing.uniroma2.it

Daniele Casali
*Department of Electronic Engineering*
*University of Rome Tor Vergata*
Rome, Italy
daniele.casali@uniroma2.it

Giovanni Costantini
*Department of Electronic Engineering*
*University of Rome Tor Vergata*
Rome, Italy
costantini@uniroma2.it

***Abstract*— Achieving advancements in automatic recognition of emotions that music can induce require considering multiplicity and simultaneity of emotions. Comparison of different machine learning algorithms performing multilabel and multiclass classification is the core of our work. The study analyzes the implementation of the Geneva Emotional Music Scale 9 in the Emotify music dataset and investigate its adoption from a machine-learning perspective. We approach the scenario of emotions expression/induction through music as a multilabel and multiclass problem, where multiple emotion labels can be adopted for the same music track by each annotator (multilabel), and each emotion can be identified or not in the music (multiclass). The aim is the automatic recognition of induced emotions through music.**

## I. INTRODUCTION

Music has the power of inducing emotions, and human beings exploit such a phenomenon in order to empower a variety of mental states and activities, both positively and negatively. The study of emotions and music has a long and still vibrant tradition. More recent is the field investigating music emotion recognition through computational means. Music emotion recognition (MER) is an emerging and cross-disciplinary field spanning information retrieval (audio, symbolic and metadata) and machine learning, on a strong backing of music cognition (semiology of music and psychology) and music theory. Musical stimuli can be categorized according to the emotions that they can induce. As computational means have progressively increased efficiency and provided more accurate results, the contribution of MER to the general field of emotion research has become central in the study of the emotional expressiveness of music. The automatic recognition of the emotions that music can induce through listening is an important part of automated music information retrieval (MIR). Computational approaches for MER through automated retrieval of musical information have shown accuracy that is comparable or surpasses human performance [1]. Emotion induction through music is the process of emotionally affecting a subject through musical stimuli. However, the complexity of human emotions is still neither fully understood from a theoretical perspective, nor fully captured and represented using computational means. Approaches narrowing down the complexity of the phenomenon from specific perspectives are known for the biological, the cognitive and the cultural [2-8]. However, such implications also suggest that perception is multimodal and music stimuli can channel one single emotion or multiple emotions simultaneously. Studies approaching MER and MIR multimodally as well as emotions in their multiplicity are rare. Research in the Musical-Moods project encompasses a cross-modal approach to machine learning for validating a MER computational model. This project investigates music from a multimodal perspective that involve motor, kinesthetic, visual and language besides auditory components, and evaluating results through creative practice with the aim to understand how we feel and attribute meaning when interacting with music technology. The Musical-Moods [9] dataset targets emotions and mental states indexed by language modelling of the participants and comprises audio excerpts, vector-based 3D animations and dance video recordings from automatic music generation through an interactive and music system [10] and professional dancers. In the present paper, we want to identify the best methods to categorize and recognize simultaneous music-induced emotions, in order to know what to expect in the Musical-Moods project before the dataset is produced and use these best methods in the definition of the MIR component of its computational model.

There are multiple ways to categorize emotions and music in terms of emotions. A first and simple approach is to have a listener indicate what is the emotion expressed by a set of music stimuli. Difficulties in emotion representation can be encountered in terms of annotator's subjectivity and language constraints. Moreover, different subjects can have a different opinion about the same emotional content. The need for a universal model able to represents real-world scenarios is paramount. An important first approach is to consider multiple annotation to describe the same music for each music piece of a dataset as different annotators can annotate a subset of the dataset through predefined tags. This known as multiclass annotation [11,14]. A more nuanced approach capable of better modelling emotion induction is to use continuous values representing different dimensions of the perceived emotions which are attributed to a specific piece [15,16], typically adopting the valence-arousal Cartesian expressive space [1]. These dimensions are independent, with valence indicating how a person feels depending on positive or negative evaluations of people, things or events, and arousal indicating the degree of a person's activation and his/her inclination to perform actions. It is possible to extend the valence-arousal space by a third dimension as discussed in MIR-based psychology [4]. A more sophisticated approach is to build a dictionary of discernible patterns appearing in sample data on a short timescale and then expressing the contents of a music file using contents of the dictionary. We will focus our study on the latest implementation of this latter approach.

## II. Materials and Methods

### A. Emotify database and GEMS

For testing our system, we use the Emotify music dataset [17]. It includes 400 tracks (44100 Hz, 128 kbps, one minute each) and incorporates four music genres (classical, rock, pop, electronic music), 100 tracks per genre and 8407 annotations total.

Annotations in the dataset are collected using the Genevan Emotional Music Scales (GEMS) [11], that consists of 45 terms for annotation of musically induced/expressed emotions. Shorter versions of 25 and 9 terms exist [12]. This can be considered as an emotional cluster model [19], investigating array of terms for music related emotions that describes the emotional and psychological response of music annotators. Each annotator could select maximally three items from the scale in order to describe the emotions which they felt most prominent when listening to a music track from the dataset. In the dataset, 515 terms are noted for the verbal description of the emotions induced in the listening. The different arrays of terms are then grouped on the results of a set of experiments with different annotators, as reported below.

Experiment 1 and 2 (Total Participants: 354): a list of terms related to emotions that are relevant to the music is made. Attention is also paid to perceived and induced emotions through 5 groups of listeners with distinct musical preferences. Experiment 3 (Total participants: 801): the structure of emotions for music induced by Cronbach's alpha factor is examined. You get a model with 9 emotions factors for induced music. Experiment 4 (Total Participants: 238): the model in Experiment 3 is replicated and it is found that it better represents the emotions needed to describe the psychology of listening to music.

In the present study, we investigate the adoption of GEMS from a machine-learning perspective. The aim is the automatic classification of induced emotions through music using a multilabel approach. We introduce a new method for feature extraction, that integrates different frameworks and best methods for automatic feature selection and grid search for the optimization of the classification parameters.

The annotations in Emotify dataset uses a version of the scale representing nine emotions: GEMS-9. The annotations, herein referred to as labels, verify the inducement of an emotion among the following: amazement, solemnity, tenderness, nostalgia, calmness, power, joyful activation, tension, sadness. The label identifies the class a track belongs to, through statistical evaluation.

### B. Threshold in the annotation

At a first glance, it is evident that the amount of total annotations differs for each of the 400 tracks in the corpus. The simplest statistical criterion that we can use to describe the distribution of the annotations is to compute the mean positive response of the total annotations. This method provides information regarding the inducement of a specific emotion in the annotators through the music tracks. Because the dataset adopts a multiclass approach, the probability of data sparseness is high when considering the expected mean annotation of a single emotion per each track (2.335), meaning that track can have labels produced by few annotations. We consider a label to be valid if the mean positive response of annotations exceeds a specific threshold, herein defined as consensus threshold. The method produces a score value for each label which represents that emotion in the track. This score is also computed in [11], as capable of preserving information regarding the emotional expressiveness and emotion induction in listening for each track.

$$score_{ij} = \frac{1}{n_j} \sum_{k=1}^{n_j} a_k^{ij} \qquad (1)$$

where:

- $i$ is the i-th emotion
- $j$ is the j-th track
- $n_j$ is the total number of annotations for the j-th track
- $a_k^{ij}$ is the presence or absence (1 or 0) of the i-th emotion in the j-th track, expressed according to a specific annotator $k$.

The method computes the score without weighing the number of listening sessions for each annotator.

Advancements in MER can be achieved by generalizations capable of surpassing the intrinsic difficulties existing in producing a universal model of emotion induction. We analyze further the Emotify music corpus and the implementation of the GEMS-based multiclass annotation approach by adopting the consensus threshold technique as the criteria for considering an emotion to be relevant for a certain track. The emotion score is computed as defined in (1), in order to consider a label as valid if exceeding a defined threshold. In this way, a score represents a consistent emotional expressiveness for emotion induction in listening among the annotators regarding a single emotion.

We estimate the percentage consensus threshold through a posteriori observation of the average trend of the emotions per track in the dataset. The percentage value represents the proportion of annotators who identify specific emotions as part of the average trend. This average function is monotonically decreasing, and its standard deviation has a decreasing trend that is stable to the unit value. Notably, we identify a plateau zone at 30%. The corresponding estimated thresholds have an average of 2.5 ± 1 annotated mood per track

### C. Feature Extraction

We extract audio features from the audio files of the Emotify dataset by using different computational frameworks such as MIRToolbox and Marsyas for deriving statistical functions, and Psysound for extracting psychoacoustical features. The extracted features can be grouped into four classes:

1) *Acoustic Features*
   a) *related to intensity*, such as RMS, RMS standard deviation, Less-than-average RMS
   b) *related to rhythm*, such as Rhytmic Fluctuation, Tempo, Strength of Strongest Beat, Beat Sum
   c) *related to timbre*, such as Attack Time, Attack Slope, Zero Crossing Rate, Zero Cross Derivative, Spectral Roll Off, High Frequency Energy,

Spectral Flux, Mel-Frequency Cepstral Coefficients, Roughness, Irregularity

2) *Psychoacoustic Features*

Loudness, Sharpness, Timbral Width, Volume, Virtual Pitch, Pure Tonalness, Multiplicity

3) *Melodic Features*

Pitch Salience and Duration, Pitch Mean, Pitch std, Salience Total, Salience Mean, Salience std, Contour Length, Pitch range, Salience Range, Tremolo, Vibrato

4) *Statistic Features*

features calculated from most of the previously defined features.

### D. Feature Discretization

When classifying continuous data, a variable amount of discretization error is always present. Reducing this amount to a level that can be considered as negligible in relation to the music emotion modelling, before the classification occurs, is important. To discretize the values of features, we use Kononenko's discretization [20], a process that provides optimal results for audio-based tasks. It is a recursive algorithm, whose stopping criterion is based on the Minimum Description Length principle (MDL), that considers regularities in a given set of data to compress the data and describe it through fewer symbols than in their original representation.

### E. Feature Selection

While the extraction of audio features is a major step in audio classification tasks, operating a reasoned selection of features can strongly improve the results of the classification. An excessively large number of features could lead to various disadvantages, such as increased processing time, reduced classification accuracy and information redundancy. In the training phase of the classification task, the use of too many features may expose to the risk of data overfitting. In order to select the most relevant features from a large feature set, the automatic selection is highly advisable and various algorithms exist for the automatic selection of most-relevant features. In the present article, we use the CFS (Correlation-based Feature Selection) algorithm [21], that automatically selects features with a strong correlation to the primary class and a weak correlation to the other features. This way, selected features have high relevance and low redundancy [22,23].

### F. Classification

Extracted features are part of the raw data that is used to classify tracks into nine distinct emotions (or classes). As mentioned, emotion multiplicity as approached through GEMS and Emotify (GEMS-9) leads to a multi-labelling problem. From a classification perspective, we represent efficiently the nine emotions as a computational problem that is both multilabel and multiclass. The nine emotion labels are not exclusive (multilabel) and each can either be present or absent for a music track (multiclass). In such a classification, each music track in the dataset is assigned to multiple emotion labels, creating multiple classification patterns that are valid at the same time. In the present study, we approach the problem by adopting three different types of classifiers: Support Vector Machine (SVM) [24,25], Bayesian Classifier [26] and Artificial Neural Networks (ANN) [27,28]. Configuration parameters include a linear Kernel Sequential Minimal Optimization (SMO) for SVM, a simple estimator (search algorithm k2) for the Bayesian classifier, and 50 neurons in a single hidden layer with 0.3 learning rate, 0.2 momentum and supervised backpropagation techniques for training for the ANN. All classifiers were trained through Weka software (Waikato Environment for Knowledge Analysis, University of Waikato, New Zealand) [29], using 10-folds cross-validation. A similar approach was successfully implemented in other studies of emotion recognition [30-33].

## III. Experiments and Results

An initial consideration must be made regarding the audio format that is used for the audio files of the Emotify dataset. This is the popular MPEG-2 Audio Layer III (mp3) format, which is a lossy format based on the concept of auditory masking and sound data compression occurring without major distinguishable difference in the listening when comparing to the uncompressed sound data. In a classification task of induced emotions through music, the sound data that is lost in the compression could be relevant for the features in two ways. Features mostly related to perception will be affected by the loss of sound data, although only minorly, because of the auditory masking phenomenon on which the compression is based. In this case, compression operates similarly to a feature selection, as information that is not relevant in the listening is discarded before the actual feature selection process occurs. Differently, every feature that is not related to perception will potentially be affected by the compression, as of the discarded sound data. As the Emotify datasets does not provide information regarding the uncompressed sound data, no comparison is possible to that. However, we should take these considerations into account when drawing our conclusions.

Extracting features using MIRToolbox accepts audio files using the wave format (wav). While lost sound data cannot be recovered by converting the files from mp3 to wave, we still need to convert them in order to carry out the feature extraction task. As audio files in the dataset are not perfectly equal in length although all close to 30 s, we consider the shortest audio file in the dataset as the minimum file length from which to extract a vector for each feature at a frame level. Average, standard deviation, asymmetry and kurtosis are calculated for each feature. We classify the dataset using a SMO-based SVM, an ANN with backpropagation, and a Naïve Bayesian classifier. The classification infrastructure is deployed by implementing our proprietary programming code for automated classification tasks. For each classification, the code recalls four cross validation folds and 20 different parameters initializations using various methods. First, we consider all extracted features; then we using only selected features through CFS; finally, we carry out a feature discretization before the feature selection. Extracted raw features are statistically synthesized determining a total of 476 features. Discretization and feature selection improve the analysis. Table 1 shows results of classification. These results are reported in terms of accuracy, defined as correctly classified instances, per emotion and threshold of 30. As mentioned, these thresholds represent the mean of annotators who identify the presence of an emotion and allows us to understand which emotions are better categorized by GEMS-9 in the Emotify dataset.

TABLE I. LIKELIHOOD VALUES FOR KNOWN SPEAKERS OBTAINED BY MEANS OF GMM METHOD

| | All features | | | Selected features (CFS) | | | Discretization + CFS | | |
|---|---|---|---|---|---|---|---|---|---|
| **Emotion** | **SMO** | **Bayes** | **ANN** | **SMO** | **Bayes** | **ANN** | **SMO** | **Bayes** | **ANN** |
| **Amazement** | 91.60 | 90.05 | 89.50 | 92.45 | 92.05 | 96.25 | 96.65 | 95.45 | 97.00 |
| **Calmness** | 76.75 | 76.55 | 76.00 | 80.05 | 81.95 | 74.90 | 72.25 | 83.70 | 74.80 |
| **Joyful** | 84.35 | 79.95 | 83.00 | 87.65 | 87.70 | 87.20 | 76.85 | 87.75 | 83.45 |
| **Nostalgia** | 72.10 | 74.30 | 72.00 | 74.85 | 80.35 | 71.45 | 75.10 | 81.80 | 77.95 |
| **Power** | 86.20 | 87.55 | 88.50 | 91.20 | 91.45 | 89.95 | 91.85 | 95.35 | 92.40 |
| **Sadness** | 79.20 | 74.45 | 80.50 | 81.70 | 80.25 | 83.35 | 85.60 | 89.15 | 86.70 |
| **Solemnity** | 65.25 | 65.15 | 67.00 | 69.05 | 70.60 | 65.40 | 78.10 | 80.75 | 77.55 |
| **Tenderness** | 74.50 | 71.75 | 74.50 | 75.40 | 75.60 | 72.65 | 87.90 | 89.65 | 90.50 |
| **Tension** | 67.05 | 65.95 | 67.50 | 77.05 | 74.90 | 76.70 | 84.55 | 88.40 | 84.80 |
| **Classification mean** | 77.44 | 76.19 | 77.61 | 81.04 | 81.65 | 79.76 | 83.21 | 88.00 | 85.02 |
| **Classification standard deviation** | 8.32 | 8.09 | 7.86 | 7.52 | 7.06 | 9.44 | 7.73 | 4.98 | 7.02 |

TABLE II. COMPARISON BETWEEN RMSE AND THE PROPOSED CLASSIFICATION METHODS FOR THE EMOTIFY DATASET

| | **MIRToolbox + PsySound** | **OpenSmile** | **MP + Harm** | **Musicological** | **Bayes** | **SMO** | **ANN** |
|---|---|---|---|---|---|---|---|
| | *RMSE* | *RMSE* | *RMSE* | *RMSE* | *RMSE* | *RMSE* | *RMSE* |
| **Amazement** | .99 ± .16 | .95 ± .13 | 1.05 ± .11 | .85 ± .24 | .17 ± .10 | .15 ± .11 | .12 ± .12 |
| **Calmness** | .83 ± .09 | .89 ± .07 | .78 ± .09 | .70 ± .16 | .36 ± .07 | .52 ± .07 | .45 ± .14 |
| **Joyful** | .77 ± .11 | .80 ± .08 | .75 ± .11 | .58 ± .15 | .31 ± .09 | .47 ± .08 | .35 ± .16 |
| **Nostalgia** | .82 ± .12 | .89 ± .07 | .88 ± .10 | .69 ± .16 | .39 ± .07 | .49 ± .08 | .42 ± .14 |
| **Power** | .82 ± .13 | .84 ± .09 | .80 ± .16 | .78 ± .26 | .19 ± .09 | .27 ± .09 | .19 ± .14 |
| **Sadness** | .87 ± .11 | .96 ± .18 | .88 ± .12 | .93 ± .20 | .28 ± .07 | .37 ± .07 | .29 ± .14 |
| **Solemnity** | .80 ± .09 | .95 ± .13 | .89 ± .15 | .84 ± .22 | .36 ± .07 | .46 ± .08 | .42 ± .14 |
| **Tenderness** | .84 ± .10 | .95 ± .07 | .85 ± .18 | .50 ± .19 | .26 ± .08 | .34 ± .07 | .22 ± .13 |
| **Tension** | .87 ± .20 | .94 ± .19 | .85 ± .13 | .71 ± .36 | .30 ± .07 | .39 ± .08 | .32 ± .14 |

As expected, MLP using a single hidden layer provides the most accurate classification when considering all features, although with an accuracy (77.61%) that is only slightly superior to the results obtained on the same feature set using SVM (77.44%). For all the classifiers, we achieve an improvement using CFS and running a discretization task before the feature selection. Using discretization, the highest accuracy is obtained with the Bayesian classifier and is attested at 88%. Notably, the difference in accuracy regarding the recognition of the specific emotions is consistent to previous research on the Emotify dataset. In that, all the feature sets demonstrate the same pattern of success and failure; we take that consistent classification results are reported across the different feature sets. This leads the authors to the conclusion that amazement and joyful activation must be emotional categories, which are very different in their subjectiveness. Our results, however, do not avail these conclusions, as we report high accuracy values at around 90% for both emotions. On the contrary, we obtain the lowest accuracy for the classification of the solemnity label. These effects are mainly due to the personality and culture of the individual annotator. In fact, there are some annotators who are very inclined to admit the presence of several moods in a music piece, and others who present a higher emotional threshold of acceptance. The first ones contribute to the presence of moods with low recognition, while the second ones contribute to increasing the recognition of the most perceived moods.

Notably, a direct comparison is possible by considering the Root Mean Square Error (RMSE) and its standard deviation across the multiple cross validation rounds. Table II compares the RMSE and its standard deviation with the accuracy of the proposed classification method, for each classifier and the combined of discretization and CFS. The comparison shows that all our features set outperform considerably those adopted by the authors', in all classification cases and for all mood labels.

## IV. CONCLUSIONS

Music emotion induction is a phenomenon whose complexity is simplified in most approaches to the automatic recognition of emotion in music: a listener can perceive multiple emotions simultaneously and subjective and cultural differences can constitute a bias in annotations tasks. In the present paper, we investigate classification of multiple and simultaneous emotions that can be induced/expressed through individual music tracks, with the aim of identifying best methods for information retrieval and computational learning models of emotions in music. We test a variety of approaches on the Emotify dataset, which adopts the GEMS-9 model for the categorization of nine different labels of emotion that can be expressed/induced simultaneously in music by a same annotator. The analysis of Emotify and GEMS-9 will provide us the opportunity to design better a new multimodal dataset for MER, MIR and computational creativity.

We approach the scenario of emotions expression/induction through music as a multilabel and multiclass problem, where multiple emotion labels can be adopted for the same music track by each annotator (multilabel), and each emotion can be identified or not in the music (multiclass). We consider this approach as better approximation to real-world scenarios in comparison to the use of exclusive labels for the description of emotions in music. We consider different distributions of annotations and emotion labels in a corpus by considering emotion labels as valid for a track when the mean positive response of annotations per label surpasses a specific consensus threshold. Thresholds at 30% is considered. Furthermore, we compare the efficiency of different approaches for discretization, feature selection and classification.

Best performance of classification tasks is achieved by using a Naïve Bayesian classifier at 88% mean accuracy. Results from the different classification algorithms confirm that pre-processing techniques, discretization and selection of features, improve performance in terms of accuracy.

## References

[1] U. Schimmack and R. Reisenzein, "Experiencing Activation: Energetic Arousal and Tense Trousal are Not Mixtures of Valence and Activation", *Emotion*, vol. 2, no. 4, p. 412, 2002.

[2] Sloboda, J. (2005), "Exploring the musical mind: Cognition, emotion, ability, function", *Oxford University Press*.

[3] Barthet, M., Fazekas, G., & Sandler, M. (2012, June). Music emotion recognition: From content-to context-based models. In *International Symposium on Computer Music Modeling and Retrieval* (pp. 228-252). Springer, Berlin, Heidelberg.

[4] Zentner, M., Grandjean, D., & Scherer, K. R. (2008). Emotions evoked by the sound of music: characterization, classification, and measurement. *Emotion*, *8*(4), 494.

[5] Lartillot, Olivier, Petri Toiviainen, and Tuomas Eerola. "A Matlab toolbox for music information retrieval.", *Data analysis, machine learning and applications. Springer Berlin Heidelberg*, 2008. 261-268.

[6] Turnbull, Douglas R., et al. "Combining audio content and social context for semantic music discovery." *Proceedings of the 32nd international ACM SIGIR conference on Research and development in information retrieval*. ACM, 2009

[7] Turnbull, Douglas, et al. "Semantic annotation and retrieval of music and sound effects", *IEEE Transactions on Audio, Speech, and Language Processing* 16.2 (2008): 467-476

[8] Lin, Yi, Xiaoou Chen, and Deshun Yang. "Exploration of Music Emotion Recognition Based on MIDI." *ISMIR. 2013*.

[9] Paolizzo, F. (2019), "The Musical-Moods project", *H2020-MCSA.GR659434*, http://musicalmoods2020.org

[10] Paolizzo, F., Johnson, C. G. (2020), "Creative Autonomy in a Simple Interactive Music System", *Journal of New Music Research*, DOI: 10.1080/09298215.2019.1709510

[11] Aljanaki A., Wiering F., Veltkamp, R.C., "Computational Modeling of Induced Emotion Using GEMS", *Proc. of the 15th International Society for Music Information Retrieval Conference*, ISMIR 2014, Taipei, Taiwan, pp.373-378, 2014.

[12] C. Laurier, O. Lartillot, T. Eerola, and P. Toiviainen: "Exploring Relationships between Audio Features and Emotion in Music," *Conference of European Society for the Cognitive Sciences of Music*, 2009.

[13] Tsoumakas, Grigorios, Ioannis Katakis, and Ioannis Vlahavas. "Mining multi-label data." Data mining and knowledge discovery handbook. Springer US, 2009. 667-685

[14] Bruni, Elia, Nam-Khanh Tran, and Marco Baroni. "Multimodal Distributional Semantics." *J. Artif. Intell. Res.(JAIR)*, 49.2014 (2014): 1-47

[15] Y. H. Yang, Y. C. Lin, Y. F. Su, and H. H. Chen, "A Regression Approach to Music Emotion Recognition," *IEEE Transactions on Audio, Speech, and Language Processing*, Vol. 16, No. 2, pp. 448-457, 2008.

[16] Y. E. Kim, E. M. Schmidt, R. Migneco, B. G. Morton, P. Richardson, J. Scott, J. A. Speck, and D. Turnbul, "Music Emotion Recognition: A State of The Art Review," in *11th International Society for Music Information Retrieval Conference (ISMIR 2010)*, 2010.

[17] Aljanaki, Anna, Frans Wiering, and Remco Veltkamp. "Collecting annotations for induced musical emotion via online game with a purpose Emotify", *Technical Report Series 2014*, UU-CS-2014-015 (2014).

[18] Downie, X. H. J. S., Cyril Laurier, and M. B. A. F. Ehmann. "The 2007 MIREX audio mood classification task: Lessons learned." *Proc. 9th Int. Conf. Music Inf. Retrieval*, 2008.

[19] Cronbach, L. J. "Coefficient Alpha and The Internal Structure of Tests", *Psychometrik*a, vol. 16, 1951.

[20] Kononenko, I., "On biases in estimating multi-valued attributes£", *Proceedings of the 14th International Joint Conference on Artificial Intelligence*, IJCAI 1995, vol. 2, pp. 1034–1040. Morgan Kaufmann Publishers Inc. (1995)

[21] Hall, M.A., "Correlation-based Feature Selection for Discrete and Numeric ClassMachine Learning", *Proc. of the Seventeenth International Conference on Machine Learning*, Morgan Kaufmann, San Francisco, CA, USA, 2000, pages 359–366

[22] Mencattini, A., Martinelli, E., Costantini, G., Todisco, M., Basile, B., Bozzali, M., et al. (2014). "Speech emotion recognition using amplitude modulation parameters and a combined feature selection procedure", *Knowledge-Based Systems, 63, 68–81.*

[23] Asci, F., Costantini, G., Di Leo, P., Zampogna, A., Ruoppolo, G., Berardelli, A., Saggio, G., & Suppa, A. (2020), "Machine-Learning Analysis of Voice Samples Recorded through Smartphones: The Combined Effect of Ageing and Gender", *Sensors*, 20(18), 5022.

[24] Platt, J. (1999), "Fast Training of Support Vector Machines Using Sequential Minimal Optimization", *Advances in Kernel Methods: Support Vector Learning*, 185–208.

[25] Costantini, G., Todisco, M., Perfetti, R., Basili, R., & Casali, D. (2010). "SVM Based Transcription System with Short-Term Memory Oriented to Polyphonic Piano Music", *Mediterranean Electrotechnical Conference (MELECON) 201*

[26] John, G., & Langley, P. (1995). "Estimating Continuous Distributions in Bayesian Classifiers". *Proceedings of the 11th Conference on Uncertainty in Artificial Intelligence.*

[27] Van Der Malsburg, C. (1986). "Frank Rosenblatt: Principles of Neurodynamics: Perceptrons and the Theory of Brain Mechanisms", *In G. Palm & A. Aertsen (A c. Di), Brain Theory (pagg. 245–248). Springer Berlin Heidelberg.*

[28] Saggio, G., Giannini, F., Todisco, M., & Costantini, G. (2011). "A data glove based sensor interface to expressively control musical processes", *Proc. of 2011 4th IEEE International Workshop on Advances in Sensors and Interfaces (IWASI) (pag. 195). https://doi.org/10.1109/IWASI.2011.6004715*

[29] Hall, M., Frank, E., Holmes, G., Pfahringer, B., Reutemann, P., & Witten, I. H. (2009). "The WEKA data mining software: An update", *ACM SIGKDD Explorations Newsletter, 11(1), 10–18*

[30] Schuller, B. - Rigoll, G. - Lang, M., "Speech Emotion Recognition Combining Acoustic Features and Linguistic Information in a Hybrid Support Vector Machine-belief Network Architecture", in *IEEE International Conference on Acoustics, Speech and Signal Processing (ICASSP)*, 1, pp. 577–580, 2004.

[31] Parada-Cabaleiro, E., Costantini, G., Batliner, A., Schmitt, M., Schuller, B.W. (2020), "DEMoS: an Italian emotional speech corpus: Elicitation methods, machine learning, and perception", *Language Resources and Evaluation, 54(2), 341-383*

[32] Parada-Cabaleiro E., Costantini G., Batliner A., Baird A., Schuller B. (2018), "Categorical vs Dimensional Perception of Italian Emotional Speech", *INTERSPEECH 2018*

[33] Parada-Cabaleiro, E., Schmitt, M., Batliner, A., Hantke, S., Costantini, G., Scherer, K., Schuller, B.W. (2018). "Identifying emotions in opera singing: Implications of adverse acoustic conditions", ISMIR 2018.